\documentclass[11pt,a4paper]{article}
\usepackage[czech]{babel}
\usepackage[cp1250]{inputenc}
\usepackage{graphicx}
\newcommand{\zav}[1]{\left(#1\right)}

\renewcommand{\title}[1]{\begin{center}{\large\bf #1}\end{center}}
\renewcommand{\author}[1]{\begin{center}#1\end{center}}
\newcommand{\afill}[2]{\vspace{-18pt}\begin{center}\small\textit{#1}\\ \textit{#2}\end{center}}
\newcommand{\Section}[1]{\vspace{12pt}\par\noindent\stepcounter{section}{\textbf{\thesection. #1}}\\[6pt]}
\renewcommand{\subsection}[1]{\vspace{6pt}\par\noindent\stepcounter{subsection}\textit{\thesubsection. #1}\\[6pt]} \renewcommand{\caption}[1]{\vspace{6pt}\par\noindent\refstepcounter{figure} {\small \textbf{Fig. \thefigure.} #1}}
\begin{document}
\thispagestyle{empty}
\title{BASIC RELATIONS FOR THE PERIOD VARIATION MODELS OF VARIABLE STARS}                        
\author{Z. Mikulášek$^a$, T. Gráf$^b$,  M. Zejda$^c$, L.-Y. Zhu$^d$, and S.-B. Qian$^d$}
\afill{$^a$Johann Palisa Observatory and Planetarium, VŠB--Technical
University~of~Ostrava, Dept. of~Theoretical Physics and
Astrophysics, Masaryk~University, Brno, Czech
Republic}{mikulas@physics.muni.cz} \afill{$^b$ Johann Palisa
Observatory and Planetarium, VŠB--Technical University~of~Ostrava,
Czech Republic}{tomas.graf@vsb.cz} \afill{$^c$Dept.~of Theoretical
Physics and Astrophysics, Masaryk~University,\\Brno,
Czech~Republic}{zejda@physics.muni.cz} \afill{$^d$National
Astronomical Observatories/Yunnan Observatory, \\ \it Chinese
Academy of Sciences, Kunming, China}{zhuly@ynao.ac.cn,\
qsb@ynao.ac.cn}

\begin{quotation}{\small\noindent \textbf{Abstract:}
Models of period variations are basic tools for period analyzes of
variable stars. We introduce phase function and instant period and
formulate basic relations and equations among them. Some simple
period models are also presented.
\\[6pt]
}
\end{quotation}
\vspace{-12pt}
%
\Section{Motivation} The majority of variable stars change
brightness (also radial velocity, magnetic field, etc.) more or less
regularly and with one period $P$. As a rule the form and amplitude
of their light curves remain constant for plenty of cycles while
their periods may change slightly for a number of reasons. The goal
of this paper is to create models of such period variations which
can then serve as basic tools for the advanced period analysis of
changes of variable objects.

\Section{The instantaneous period and the phase
function}\label{zakladka} The state of a periodically changing
variable star is described by two functions of time $t$: the
non-descending stairs-like epoch function $E(t)$ expressing the
number of cycles elapsed from the moment beginning of epochs
counting $M_0$, and the sawtooth function called the phase
$\varphi(t)$, reaching its minimum $(\varphi(t)=0)$ when the new
cycle of the variability starts and its maximum $(\varphi(t)=1)$
when the particular cycle ends. The phase is used for the
construction of so called phase curves of various values
characterizing the variable star. The \textit{instantaneous period}
for a selected epoch $P(E)$ is then the duration of this epoch. It
is useful to introduce a new time-dependent quantity --
\textit{phase function} $\vartheta(t)$, as
\begin{equation}\label{fi}
\vartheta(t)=E(t)+\varphi(t);\quad \varphi(t)={\mathop{\rm
frac}\nolimits}[\vartheta(t)];\quad E(t)={\mathop{\rm
floor}\nolimits}[\vartheta(t)],
\end{equation}
that could substitute both these `mathematically awful' functions.
The operator `frac' removes from a real number its integral part,
while `floor' rounds the quantity to the nearest lower integer.

The phase function $\vartheta(t)$ is a monotonic rising function of
time originating at the beginning of epoch counting $t=M_0$,
$\vartheta(M_0)=0$. Using $\vartheta(t)$ we are able to determine
the epoch and the phase (see Eq.\,\ref{fi}) for any time. The time
derivative of the phase function at the time $t$ is equal to the
instantaneous frequency $\dot{\vartheta}(t)=f(t)=P(t)^{-1}$, where
$P(t)$ is the instantaneous period of the variable star at the time
$t$ (firstly introduced in \cite{mik}). The phase function is then
determined by the following differential equation
\begin{equation}\label{tetka}
\frac{\mathrm{d}\vartheta(t)}{\mathrm{d}t}=\frac{1}{P(t)};
\quad\vartheta(t=M_0)=0;\quad\Rightarrow\quad
\vartheta(t)=\int_{M_0}^t{\frac{\mathrm{d}\tau}{P(\tau)}}.
\end{equation}

It is useful to also introduce the inversion function $T(\vartheta)$
to the phase function $\vartheta$ which determines the moment $T$ when
the phase function reaches the assigned value. The times of the zero
phase (usually the times of brightness minima or maxima)
$\mathit{\Theta}$ for the distinct epoch $E$ are then given by the
relation: $\mathit{\Theta}(E)=T(\vartheta=E)$.

The function $T(\vartheta)$ can be derived as the inversion function
of $\vartheta(t)$ or it can be found by solving the differential
equation:
\begin{equation}\label{antitetka}
\frac{\mathrm{d}T(\vartheta)}{\mathrm{d}\vartheta}=P(\vartheta);
\quad T(\vartheta=0)=M_0;\quad\Rightarrow\quad
T(\vartheta)=M_0+\int_{0}^\vartheta{P(\zeta)\ \mathrm{d}\zeta}.
\end{equation}

\Section{Some simple period models}\label{prommodely} Let us derive
the phase functions, their inverses and period time dependencies for
some astrophysically important polynomial period models.\vspace{3mm}

\noindent $\underline{P(t)=P_0}$\vspace{1mm}

\noindent The simplest period model that we use as a first
approximation assumes that the period of variations is constant:
$P(t)=P_0$. Then, using the equations (\ref{tetka}) and
(\ref{antitetka}), we get for its $\vartheta_1(t)$ and
$T_1(\vartheta)$
\begin{equation}
\vartheta_1(t)=\frac{t-M_0}{P_0};\quad
T_1(\vartheta)=M_0+P_0\,\vartheta_1.
\end{equation}

\noindent $\underline{\dot{P(t)}=\dot{P}_0}$\vspace{1mm}\\
\noindent Let's assume that the period is a linear function of the
time:
\begin{eqnarray}\label{exact}
&\dot{P}=\dot{P}_0;\quad
P(t)=P_0+\dot{P}_0\,(t-M_0)=P_0(1+\dot{P}_0\vartheta_1),
\\
&\displaystyle \frac{\mathrm{d}P}{\mathrm{d}t}
=\frac{\mathrm{d}P}{\mathrm{d}T}=\frac{\mathrm{d}P}
{\mathrm{d}\vartheta}\frac{\mathrm{d}\vartheta} {\mathrm{d}T}=
\frac{\mathrm{d}P} {\mathrm{d}\vartheta}\frac{1}{P}=\dot{P}_0;\quad
P(\vartheta)=P_0\,e^{\dot{P}_0\,\vartheta};
\label{linper}\\
&\displaystyle\vartheta(t)=\frac{1}{\dot{P}_0}\ln(1+\dot{P}_0\,
\vartheta_1);\quad T(\vartheta)=M_0+\frac{P_0}{\dot{P}_0}
\zav{e^{\dot{P}_0\vartheta}-1}.
\end{eqnarray}
Since the instantaneous rate of period $\dot{P}$ is as a rule very
slow, we can replace the real phase functions $\vartheta(t)$ and
their inverses by their Maclaurin decomposition:
\begin{eqnarray}
\vartheta(t)&=&\frac{1}{\dot{P}_0}\ln(1+
\dot{P}_0\,\vartheta_1)\doteq
\vartheta_1-\dot{P}_0\,\frac{\vartheta_1^2}{2}+
\dot{P}_0^2\frac{\vartheta_1^3}{3}\,-\dots; \label{laurin1}\\
\displaystyle T(\vartheta)&=& M_0+\frac{P_0}{\dot{P}_0}
\zav{e^{\dot{P}_0\vartheta}\!-\!1}\doteq
M_0+P_0\zav{\vartheta+\dot{P}_0\frac{\vartheta^2}{2!}+
\dot{P}_0^2\frac{\vartheta^3}{3!}\ldots} .\label{laurin3}
\end{eqnarray}
For the majority of variable stars with an inconstant period it is
valid that $\dot{P}\,\Delta t\ll P$, where $\Delta t$ is the whole
duration of its observation. The last decomposition terms in (see
Eqs. \ref{laurin1}, \ref{laurin3}) can then be neglected.
\begin{eqnarray}\label{parabola}
& \displaystyle \vartheta(t)\doteq
\frac{t-M_0}{P_0}-\frac{\dot{P_0}}{2}\zav{\frac{t-M_0}{P_0}}^2;\quad
T(\vartheta)\doteq
M_0+P_0\vartheta+P_0\dot{P}_0\frac{\vartheta^2}{2!};\nonumber \\
&P(t)=P_0+\dot{P}_0\,(t-M_0);\quad
P(\vartheta)=P_0\zav{1+\dot{P}_0\,\vartheta}.
\end{eqnarray}
The phase function and its inverse are quadratic functions and to
describe then we just need three parameters:
$M_0,\,P_0,\,\dot{P}_0$.

The exact same relations as given in Eqs. (\ref{parabola}) can be
obtained if we assume other laws for period development, such as
$P\,\dot{P}=P_0\,\dot{P}_0$ or $\dot{P}/P=\dot{P}_0/P_0$, or if we
assume the parabolic course of the phase function $\vartheta(t)$ or
its inverse $T(\vartheta)$, $T(E)$. \vspace{3mm}

\noindent $\underline{\ddot{P}=\ddot{P}_0}$\vspace{1mm}

\noindent If we find that the dependence of $\vartheta(t)$ have
apparent cubic or higher terms, we acknowledge that the first
derivative of the period is inconstant. Then the simplest period
model assumes the period to be a quadratic function of time:
\begin{equation}
P(t)=P_0+P_0\,\dot{P}_0\,\vartheta_1+
P_0^2\ddot{P}_0\,\frac{\vartheta_1^2}{2};\quad
\dot{P}(t)=\frac{\mathrm{d}P}{\mathrm{d}\vartheta_1}
\frac{\mathrm{d}\vartheta_1}{\mathrm{d}t}=\dot{P}_0+
P_0\,\ddot{P}_0\,\vartheta_1.
\end{equation}
Using Eq. (\ref{tetka}) we can find the exact solution for the phase
function $\vartheta(t)$. However, the function expression is so
complex that it is practically useless. Consequently we only the
first three terms of its Mclaurin decomposition.
\begin{eqnarray}
& \displaystyle \vartheta(t)\doteq \vartheta_1-
\dot{P}_0\frac{\vartheta_1^2}{2!}\,-(P_0\,\ddot{P}_0-2\dot{P}_0^2)
\frac{\vartheta_1^3}{3!}\doteq  \vartheta_1-
\dot{P}_0\frac{\vartheta_1^2}{2!}\,-P_0\,\ddot{P}_0
\frac{\vartheta_1^3}{3!};\label{th3}\\
& \displaystyle T(\vartheta)\doteq
M_0+P_0\,\vartheta+P_0\,\dot{P}_0\,\frac{\vartheta^2}{2!}+
(P_0^2\,\ddot{P}_0+P_0\,\dot{P}_0^2) \frac{\vartheta^3}{3!}\nonumber \\
& \displaystyle \doteq
M_0+P_0\,\vartheta+P_0\,\dot{P}_0\,\frac{\vartheta^2}{2!}+
P_0^2\,\ddot{P}_0 \frac{\vartheta^3}{3!};\quad \ \ \label{T3}\\
& \displaystyle \frac{P(\vartheta)}{P_0}\doteq 1
+\dot{P}_0\,\vartheta+(P_0\,\ddot{P}_0
+\dot{P}_0^2)\frac{\vartheta^2}{2!}\doteq 1
+\dot{P}_0\,\vartheta+P_0\,\ddot{P}_0
\frac{\vartheta^2}{2!}.\label{P3}
\end{eqnarray}
The expressions in brackets in Eqs. (\ref{th3}),(\ref{T3}), and
(\ref{P3}) can be simplified, because of $P_0\ddot{P}_0 \gg
\dot{P}_0^2$ is valid for all known cases of variable stars changing
their period.

\Section{Conslusion} We introduced several simple polynomial period
models that can generally be used for most known and unknown period
trends. More complicated models of period changes in eclipsing
binary systems, accounting for the light time effect caused by third
bodies as well as apsidal motion, should also be developed using the
relations given in Section 2. However, the formulation of those
period models is beyond the scope of this article.

\vspace{12pt} \noindent
\textbf{Acknowledgements}\\[6pt]
We thank Stefanus de Villiers for his kind reading of the manuscript
and correcting the language. This work was supported by LH12175, and
the intergovernmental cooperation project between P.\,R. China and the
Czech Republic.

\renewcommand{\refname}{
\end{document}